\documentclass{article}

\usepackage{subfigure}
\usepackage{graphicx}
\usepackage{xcolor}
\usepackage{geometry}
\usepackage{latexsym}
\usepackage{caption}
\usepackage{amsmath}
\usepackage{amssymb}
\usepackage[colorlinks=true,linkcolor=blue,citecolor=blue]{hyperref}
\usepackage{amsfonts}
\usepackage{algorithm, algpseudocode}
\usepackage{comment}
\usepackage{version}%
\providecommand{\U}[1]{\protect\rule{.1in}{.1in}}
\definecolor{royalazure}{rgb}{0.0, 0.22, 0.66}
\definecolor{royalblue(traditional)}{rgb}{0.0, 0.14, 0.4}
\definecolor{cerisepink}{rgb}{0.93, 0.23, 0.51}
\newtheorem{thm}{Theorem}[section]
\newtheorem{defn}[thm]{Definition}
\newtheorem{lem}[thm]{Lemma}

\newtheorem{prop}[thm]{Proposition}

\newtheorem{claim}{Claim}
\begin{document}

\title{A greedy approximation algorithm for the minimum $(2,2)$-connected dominating
set problem}
\author{Yash P. Aneja \\
Odette School of Business \\
University of Windsor \\
Windsor, Canada \\
\and Asish Mukhopadhyay \\
School of Computer Science \\
University of Windsor \\
Windsor Canada \\
\and Md. Zamilur Rahman \\
School of Computer Science \\
University of Windsor \\
Windsor Canada}

\date{}
\maketitle


\noindent{{\large \textbf{Abstract}}} Using a connected dominating set (CDS)
to serve as the virtual backbone of a wireless sensor network (WSN) is an
effective way to save energy and reduce the impact of broadcasting storms.
Since nodes may fail due to accidental damage or energy depletion, it is
desirable that the virtual backbone is fault tolerant. This could be modeled
as a $k$-connected, $m$-fold dominating set ($(k,m)$-CDS). Given a virtual
undirected network $G=(V,E),$ a subset $C\subset V$ is a $(k,m)$-CDS of $G$ if
(i) $G[C],$ the subgraph of $G$ induced by $C,$ is $k$-connected, and (ii)
each node in $V\backslash C$ has at least $m$ neighbors in $C.$ We present a
two-phase greedy algorithm for computing a $(2,2)$-CDS that achieves an
asymptotic approximation factor of $(3+\ln(\Delta+2)),$ where $\Delta$ is the
maximum degree of $G.$ This result improves on the previous best known
performance factor of $(4+\ln\Delta+2\ln(2+\ln\Delta))$ for this problem.

\section{Introduction}

Suppose $G=(V,E)$ is a connected graph. A subset $C$ of $V$ is a said to be a
connected dominating set (CDS) of $G$ if $G[C]$, the induced graph on $C$, is
connected and every vertex $v$ in $V\setminus C$ is a neighbor of $C$
(connected by an edge to some vertex $u\in C$). Nodes in $C$ are called
\textit{dominators}, and nodes in $V\backslash C$ are called
\textit{dominatees}. To save energy and reduce interference, it is desirable
that the CDS size is as small as possible. Computing a minimum CDS is a well
known NP-hard problem \cite{Garey-Johnson}. By showing that finding a minimum
set cover is a special case of finding a minimum CDS, Guha and Khullar
\cite{Guha_Khuller_1998} established that a minimum CDS can not be
approximated within $\rho\ln n$ for any 0%
$<$%
$\rho<1$ unless $NP\subset DTIME(N^{O(\log\log n)}).$ In the same paper Guha
and Khullar \cite{Guha_Khuller_1998} proposed a two-phase greedy algorithm,
with an approximation factor of $(3+\ln\Delta)$ for fining a minimum sized
CDS. Subsequently, Ruan et. al. \cite{Ruan_Du_Jia_Wu_Li_Ko_2004} used a
potential function approach to come up with a single phase greedy algorithm
improving the approximation ratio to $(2+\ln\Delta).$ There are, in the
literature, several approximation algorithms for finding a minimum CDS for a
general graph \cite{Du_et_al-2012}.

To make a virtual backbone more robust to deal with frequent node failures in
WSNs, researchers have suggested using a $(k,m)$-CDS. As mentioned in the
abstract, $C\subset V$ is a $(k,m)$-CDS if every node in $V\backslash C$ is
adjacent to at least $m$ nodes in $C,$ and $G[C],$ the subgraph induced by
$C,$ is $k$-connected. The $k$-connectedness\ means that $|C|>k$ and
$G[C\backslash X]$ is connected for any $X\subset C$ with $|X|<k$. In other
words, no two vertices of $G[C]$ are separated by removal of fewer than $k$
other vertices of $C$. With such a $C,$ messages can be shared by the whole
network, where every node in $V\backslash C$ can tolerate up to $m-1$ faults
(node failures) on its dominators, and the virtual backbone $G[C]$ can
tolerate up to $k-1$ faults.

Zhou et al. \cite{Zhou_Zhang_Wu_Xing_2014}, using a more complex potential
function than the one in Ruan et al. \cite{Ruan_Du_Jia_Wu_Li_Ko_2004}, provide
a single phase $(2+\ln(\Delta+m-2))$-approximation algorithm for the minimum
$(1,m)$-CDS problem in a general graph.

Shi et al. \cite{Shi_Zhang_Zhang_Wu_2016}, using a two-phase approach, provide
a $(\alpha+2(1+\ln\alpha))$-approximation algorithm for the minimum
$(2,m)$-CDS, where $m\geq2$ and $\alpha$ is the approximation ratio for the
computation of a $(1,m)$-CDS. Using the solution obtained for the minimum
$(1,m)$-CDS problem, they augment the connectivity of $G[C]$ by merging blocks
(a block is defined as a maximal connected subgraph without a cut-vertex) of
$G[C]$ recursively. When $m=2,$ this approximation ratio becomes $(4+\ln
\Delta+2\ln(2+\ln\Delta)).$

In this paper, we present a different two-phase approach to the $(2,2)$-CDS
problem. The first phase ends up obtaining a $C$ such that it is a
\textit{2-fold} dominating set, and all connected components of $G[C]$ are
biconnected ($2$-connected). The second phase, at each iteration, needs two
nodes from $V\backslash C$ to reduce the number of these biconnected
components by at least one. This results in an algorithm with an asymptotic
approximation factor of $(3+\ln(\Delta+2)).$ By a simple modification of the
potential function, our approach provides a $(3+\ln(\Delta+m))$-approximation
algorithm for computing a $(2,m)$-CDS.

For related and earlier work, the reader may refer to the papers by
\cite{Shi_Zhang_Zhang_Wu_2016} and \cite{Zhou_Zhang_Wu_Xing_2014}.

\section{Main results}

Let $G=(V,E)$ be a \emph{biconnected} graph. For a $C\subset V,$ define $p(C)$
to be the number of (connected) components of $G[C],$ the subgraph induced by
$C.$ Define $G\langle C\rangle$ to be the \textit{spanning} subgraph of $G,$
with vertex set $V,$ and edge set $\{e\in E:e$ has at least one end in $C\}.$
Let $q(C)$ represents the number of components of $G\langle C\rangle.$ For
each node $v\in V$, define $m_{C}(v)$ as:
\[
m_{C}(v)=%
\begin{cases}
0, & \text{if }v\in C,\text{or adjacent to at least }2\text{ nodes in $C$}\\
1, & \text{if }v\in V\setminus C,\text{ and adjacent to at most }1\text{ node
in }C\text{.}%
\end{cases}
\]

Let $m(C)=\sum_{v\in V}m_{C}(v).$ Thus $m(C)$ represents the number of nodes
in $V\backslash C$ which have at most one neighbor in $C.$ Note that for $m =
2$, $q(C)$ and $m(C)$ are defined exactly as in~\cite{Zhou_Zhang_Wu_Xing_2014}%
. Again, as in \cite{Zhou_Zhang_Wu_Xing_2014}, we assign a color to each node
in $V$ relative to a given $C$ as follows. All nodes in $C$ are colored black,
nodes in $V\backslash C$ which have at least two neighbors in $C$ are colored
gray, nodes in $V\backslash C$ that have exactly one neighbor in $C$ are
colored red, and all other nodes are colored white. 

Given $C$, we define $\hat{p}(C)$ to be%
\begin{equation}%
\begin{split}
\hat{p}(C)  &  =\max_{x\in C}{p(C\setminus\{x\})}\\
&  =p(C\setminus\{x_{C}\})
\end{split}
\label{P-hat}%
\end{equation}

in an attempt to capture the bi-connectivity deficit of $G[C]$.

A node $x_{C}$ for which the maximum in (\ref{P-hat}) is attained is called a
critical node of $C.$ Note that if $G[C]$ is biconnected then $\hat{p}(C)=1,$ in which
case every node in $C$ can be viewed as a critical node.

Finally, we use the functions, $\hat{p}(C), q(C)$ and $m(C)$ to define a
potential function, $f(C)$, on $C$ as:%
\begin{equation}
f(C)=\hat{p}(C)+q(C)+m(C)
\end{equation}

and the difference function $\Delta_{y}f(C)$ by
\[
\Delta_{y}f(C)=f(C)-f(C \cup\{y\}),
\]

where $y\in V.$ We can also, equivalently, write%
\[
\Delta_{y}f(C)=\Delta_{y}\hat{p}(C)+\Delta_{y}q(C)+\Delta_{y}m(C)
\]

\noindent\emph{Result}. Function $f(C)$ is monotonically non-increasing. That
is, $\Delta_{y}f(C)\geq0$ for every $v$ in $V$. We need to consider three cases:

\noindent\textbf{Proof:} Several cases arise.

\begin{enumerate}

\item Suppose $y$ is gray. This means that $\hat{p}(C\cup\{y\})\leq\hat{p}%
(C)$. Clearly $m(C\cup\{y\})\leq m(C),$ and $q(C\cup\{y\})\leq q(C).$ Thus,
$f(C\cup\{y\})\leq f(C),$ and hence $\Delta_{y}f(C)\geq0.$

\item Suppose $y$ is red. It is then connected to only one node in $C$. As $y$
is added to $C,$ its $m$-value goes down by one and its $q$-value cannot
increase. Its $\hat{p}$ value may increase by $1$. Thus, $f(C\cup\{y\})\leq
f(C)$.

\item Suppose $y$ is white. As $y$ is added to $C$, its $m$-value goes down by
one, $q$-value goes down by at least one, and $\hat{p}$-value goes up by one.
Hence, $f(C\cup\{y\})\leq f(C)$. 



\end{enumerate}

The following characterization of the structure of a biconnected graph
\cite{Diestel_2000} is useful for us.

\begin{defn}
Given a graph $H,$ we call a path $P$ an $H$-$path$ if $P$ meets $H$ exactly
in its ends.
\end{defn}

\begin{figure}[h]
\centering
\includegraphics[scale = 0.75]{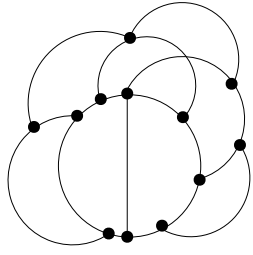}
\caption{\emph{H-path construction of 2-connected graphs}}%
\label{fig:2connectedGraphs}%
\end{figure}

For example consider a biconnected graph that is a cycle $H$ of three nodes:
$x_{1},x_{2},$ and $x_{3}.$ Then a path $P$ of three nodes $x_{1},x_{4},$ and
$x_{3}$ is an $H$-path of $H.$ Adding this $H$-path to cycle $H$, keeps it
biconnected. The following proposition formalizes this observation and is
illustrated in Fig.~\ref{fig:2connectedGraphs}

\begin{prop}
\cite{Diestel_2000} A graph $H$ is biconnected if and only if it can be
constructed from a cycle by successively adding $H$-$paths$ to graphs $H$
already constructed.
\end{prop}


\begin{figure}[h]
\centering
\includegraphics[scale =1]{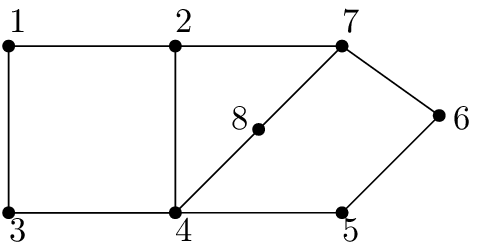} 
\caption{{}}%
\label{Fig-exampleGraph}%
\end{figure}

Suppose $C^{\ast}$ is a minimum $(2,2)$-CDS of $G.$ Since it is biconnected,
using the above proposition we can list the nodes in an order such that each
sublist starting from the beginning is essentially a \textquotedblleft%
\textit{path}", where the first node of this \textquotedblleft\textit{path}"
might correspond to a biconnected subgraph of $C^{\ast}.$ Let us illustrate
this with the following example~\ref{Fig-exampleGraph} of $G[C^{\ast}].$

We can list 8 nodes of this graph as the following list with sublists:
$((1,2,3,4),5,6,7),8).$ Node $2$ is adjacent to node $1,$ node $3$ is adjacent
to only node $1.$ Node $4,$ however, is adjacent to both nodes $2$ and node
$3.$ So $(1,2,3,4)$ corresponds to a biconnected graph (cycle), and is now
designated as a ``single meta-node" in our list. Next, the $H$-path
$(4,5,6,7,2)$ is added to this subgraph, resulting in another biconnected
subgraph. Finally, adding the $H$-path $(7,8,4)$ results in $G[C^{\ast}].$ The
next lemma exploits this interpretation of a biconnected graph as a ``path".

\begin{lem}
For any two subsets $A,B\subseteq V$ and any node $y\in V$, if $B$ is
a\textquotedblleft path" then
\begin{equation}
\Delta_{y}f(A\cup B)\leq\Delta_{y}f(A)+1
\end{equation}

\end{lem}

\noindent\textbf{Proof:}\bigskip The result is obvious if $y\in A.$ Suppose
$y\in B\backslash A.$ Then the above result follows as $\Delta_{y}f(A)\geq0$
for all $y\in V.$ Thus we assume from here on that $y\notin A\cup B.$ Define
$\mu(f)=\Delta_{y}f(A\cup B)-\Delta_{y}f(A).$ It is useful to write $\mu(f)$
as:
\[
\mu(f)=\mu(\hat{p})+\mu(q)+\mu(m).
\]
We first look at $\mu(m).$ Define $S$ to be set of nodes which are neighbors
of $y$ which are white with respect to $A$ and red with respect to $B$. Let
$|S|=s.$ We first want to show that:
\begin{equation}
\mu(m)=\left\{
\begin{array}
[c]{l}%
s-1,~\text{if }y\text{ is gray for }A\cup B,\text{ \textit{but} not gray for
}A\\
s,\text{ otherwise.}%
\end{array}
\right.
\end{equation}
It is easy to formalize and establish this result by looking at the following
two example figures: case (i): $\mu(m)=s-1,$ case (ii): $\mu(m)=s$.

\begin{figure}[tbh]
\centering
\subfigure[$\mu(m)=0$, $\mu(q)=-2$ \label{Fig-example2}]{\includegraphics[scale=1]{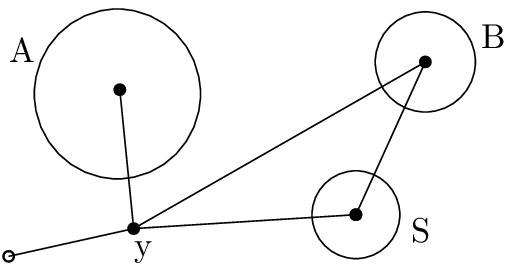}}\hfill
\subfigure[$\mu(m)=1$, $\mu(q)=-1$ \label{Fig-example3}]{\includegraphics[scale=1]{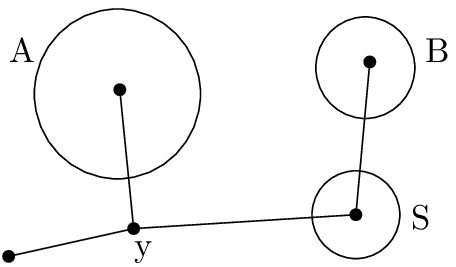}}
\caption{Two cases covering the computation of $\mu(m)$}%
\label{fig:Examples}%
\end{figure}

In both 3(a) and 3(b) of Figure~\ref{fig:Examples}, $|S|=s=1.$ In Fig.~3(a),
$m(A)=4,$ $m(A\cup\{y\})=3.$ Hence $\Delta_{y}m(A)=1.$ $m(A\cup B)=2,$
$m(A\cup B\cup\{y\})=1.$ Hence $\Delta_{y}m(A\cup B)=1.$ So $\mu(m)=\Delta
_{y}m(A\cup B)-\Delta_{y}m(A)=0.$

In Fig.~3(b), $m(A)=4,$ $m(A\cup\{y\})=3.$ Hence $\Delta_{y}m(A)=1.$ $m(A\cup
B)=3,m(A\cup B\cup\{y\})=1,$ $\Delta_{y}m(A\cup B)=2.$ Hence $\mu
(m)=\Delta_{y}m(A\cup B)-\Delta_{y}m(A)=1.$\bigskip

We now look at $\mu(q).$ We want to show that:
\begin{equation}
\mu(q)\leq\left\{
\begin{array}
[c]{l}%
-s,~\text{if }y\text{ is adjacent to }B\\
-(s-1),\text{ otherwise.}%
\end{array}
\right.
\end{equation}

Let $N_{A}(y)$ be the set of components of $G\langle A\rangle$ that are
adjacent with node $y$ in $G$ (the component of $G\langle A\rangle$ containing
node $y$, if any, is not counted). Then $\Delta_{y}q(A)=|N_{A}(y)|.$ Hence
$\mu(q)=|N_{A\cup B}(y)|-|N_{A}(y)|.$ Again, it is easy to formalize and
establish the above result by looking at the above two example figures, in
Figure 3,
covering the cases: 3(a), $\mu(q)\leq-s,$ and 3(b): $\mu(q)\leq-(s-1)$.

In Figure 3(a), $\Delta_{y}q(A)=|N_{A}(y)|=3,$ $\Delta_{y}q(A\cup B)=N_{A\cup
B}(y)=1.$ Hence $\mu(q)=1-3=-2.$

In Figure 3(b), $\Delta_{y}q(A)=N_{A}(y)=2,$ $\Delta_{y}q(A\cup B)=|N_{A\cup
B}(y)|=1.$ Hence $\mu(q)=-1.$ Zhou et al. \cite{Zhou_Zhang_Wu_Xing_2014} have
established the above two results in a more general setting.

Now we focus on $\mu(\hat{p})=\Delta_{y}\hat{p}(A\cup B)-\Delta_{y}\hat{p}%
(A)$. 

Let $r$ be a critical node of $G[A]$. Let $A_{r}$ be the set of nodes in the
component of $G[A]$ containing node $r.$ [Note that if $G[A]$ is connected
then $A_{r}=A.$]

We define three constants $\alpha$, $\beta$, and $\gamma$ in $G[A]$ as
follows. Let

$\alpha$ = $p(A_{r}\backslash\{r\}),$ the number of components in
$G[A_{r}\backslash\{r\}].$ 

$\beta$ = The number of components $G[A\backslash A_{r}]$ which are adjacent
to node $y$ in $G[A]$.

$\gamma$ = The number of components of $G[A_{r}\backslash\{r\}]$ which are are
adjacent to node $y$ in $G[A].~(\gamma\leq\alpha)$.

\begin{figure}[tbh]
\centering
\includegraphics[scale=1]{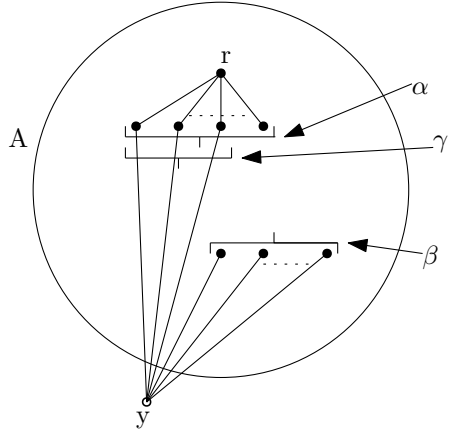}
\caption{Illustrating the parameters $\alpha$, $\beta$ and $\gamma$}%
\label{Fig-alphaBetaGamma}%
\end{figure}

Refer to Fig.~\ref{Fig-alphaBetaGamma} for an illustration.

\noindent\textit{Result-1:} $\Delta_{y}\hat{p}(A) = \min\{\alpha, \beta
+\gamma\}-1$.

\noindent\textbf{Proof:} Referring to the figure above, note that $\hat
{p}(A)=\alpha+\beta$. Now let us calculate $\hat{p}(A\cup\{y\})$. Whichever of
the two nodes, node $r$ or node $y,$ whose removal results in the higher
number of components in $G[A\cup\{y\}]$ is the critical node . Now if we
remove node $r$, the resulting number of components will be $(\alpha
-\gamma)+1$. If we remove node $y$ then this number is $\beta+1$. Hence
\[
\hat{p}(A\cup\{y\})=\max\{\alpha-\gamma,\beta\}+1
\]
Hence,
\[%
\begin{split}
\Delta_{y}\hat{p}(A)  &  =\hat{p}(A)-\hat{p}(A\cup\{y\})\\
&  =\alpha+\beta-[\max\{\alpha-\gamma,\beta\}]+1\\
&  =\alpha+\beta+\min\{\gamma-\alpha,-\beta\}-1\\
&  =\min\{\beta+\gamma,\alpha\}-1
\end{split}
\]

Returning to $\mu(\hat{p})=\Delta_{y}\hat{p}(A\cup B)-\Delta_{y}\hat{p}(A)$,
we use \textit{result-1} to make some assertions about $\mu(\hat{p})$. As we
mentioned earlier, we can assume that $B$ is a set of nodes which form a
\textquotedblleft path". Since $B$ is a \textquotedblleft path", adding $B$ to
$A$ does not create a new critical node in $G[A\cup B]$. 

\noindent\textit{Result-2:} Suppose $y$ is not adjacent to $B$, then $\mu
(\hat{p})=0$.

\noindent\textbf{Proof:} Since $y$ is not adjacent to $B$, adding $B$ to $A$
does not change $\beta$ and $\gamma$ values. $\alpha$ value may increase.
Hence, $\min\{\beta+\gamma,\alpha\}$ does not change, implying $\mu(\hat
{p})=0$. 

\noindent\textit{Result-3:} If $y$ is adjacent to $B$, then $\mu(\hat{p}%
)\leq1$.

\noindent\textbf{Proof:} If $B$ is not adjacent to $r$, then $\beta$ goes up
by $1$, $\alpha$ and $\gamma$ do not change. Hence $\mu\leq1$. If $B$ is
adjacent to $r$, then both $\alpha$ and $\gamma$ go up by $1$, but $\beta$
does not change. Hence $\mu(\hat{p})=1$. Hence we have the third \ inequality:%
\begin{equation}
\mu(\hat{p})\text{~~}\left\{
\begin{array}
[c]{l}%
=0,\text{ if }y\text{ is not adjacent to }B,\\
\leq1,\text{ otherwise.}%
\end{array}
\right.
\end{equation}
\newline

Combining the three inequalities (4), (5), and (6), proves our Lemma 2.3.




\begin{lem}
Let $G=(V,E)$ be a biconnected graph. Then, $C$ is a $2$-fold dominating set
if $\Delta_{y}f(C)=0$ for every $y\in V$.
\end{lem}

\noindent\textbf{Proof:} The following claims establish the proof.

\noindent\textit{Claim-1.} $C\neq\emptyset$.

Suppose $C=\emptyset$. We have $\hat{p}(\emptyset)=0,~q(\emptyset
)=|V|,~m(\emptyset)=|V|.$ Since $G$ is biconnected, every node in $G$ has
degree at least 2. Pick any node $y.$ So $C=\{y\},$ and $\hat{p}(\{y\})=0,$
$q(\{y\})=|V|-|N_{G}(y)|,$ $m(\{y\})=|V|.$ Hence $\Delta_{y}f(C)>0,$ a contradiction.

\noindent\textit{Claim-2. }$|C|\geq3.$ Its proof is straightforward.

\noindent\textit{Claim-3. }$m(C)=0.$ This claim would imply that $C$ is a
2-fold CDS.

Suppose $m(C)>0.$ This means that there is at least one node $y$ which is red
or white with respect to $C.$ Suppose that $y$ is a white node. This means
that $y$ is an isolated node in $G\langle C\rangle,$ and hence accounts for
one component in computing $q(C).$ Adding $y$ to $C$ implies $\Delta
_{y}q(C)\geq1,$ and $\Delta_{y}m(C)=1.$ Since $\Delta_{y}\hat{p}(C)\geq-1,$ we
have $\Delta_{y}f(C)\geq1,$ a contradiction. So assume that there are no white
nodes. Suppose $y$ is red. This mean that $y$ is adjacent to only one node in
$C.$ Since $G$ is biconnected, $y$ is adjacent to another node $y_{1}\notin
C.$ So $y_{1}$ is either red or gray. Suppose $y_{1}$ is red. Adding $y$ to
$C$ makes $y_{1}$ gray. Hence $\Delta_{y}m(C)=2.$ Since $\Delta_{y}q(C)\geq0,$
and $\Delta_{y}\hat{p}(C)\geq-1,$ we have $\Delta_{y}f(C)\geq1,$ a
contradiction. So assume $y_{1}$ is gray. Then $\Delta_{y_{1}}\hat{p}%
(C)\geq0,$ $\Delta_{y_{1}}m(C)=1,$ since $\ y$ is now gray in $G[C\cup
\{y_{1}\}]$, $\Delta_{y_{1}}q(C)\geq0,$ implying $\Delta_{y_{1}}f(C)>0,$ a
contradiction. This proves the claim.

\noindent\textit{Claim-4}. Every (connected) component in $G[C]$ is biconnected.

To prove this, suppose $C_{1}$ is a component of $C$ which is not biconnected.
Hence $C_{1}$ has a critical vertex $x$ such that $\hat{p}(C_{1}%
)=p(C_{1}\setminus\{x\})=t\geq2$. Since $G$ is biconnected, there exists a
gray node $y$ that is connected to two different components of $G[C_{1}%
\setminus\{x\}]$. Hence $\hat{p}(C_{1}\cup\{y\})\leq t-1$, implying
$\Delta_{y}\hat{p}(C)\geq1$, $\Delta_{y}q(C)\geq0$, and hence $\Delta
_{y}f(C)>0$, a contradiction. 



When $\Delta_{y}f(C)=0$, $\forall$ $y\in V\setminus C$, we say that phase-I of
the algorithm has ended.
A formal description of the Phase I algorithm is given below.
At the end of Phase I, $G[C]$ has $t$ biconnected components, $t\geq1.$ If
$t=1,$ there is nothing more to do. Again, since $G$ is biconnected, if
$C_{1}$ and $C_{2}$ are any two components 
of $G[C]$, there must exist at least two nodes $y_{1}$ and $y_{2}$ in
$V\setminus C$ such that both $y_{1}$ and $y_{2}$ are connected to both
$C_{1}$ and $C_{2}$, making $G[C\cup\{y_{1},y_{2}\}]$ having one less
component than $G[C]$.

So, if at the end of phase-I, we have $t$ components in $G[C]$, we need to add
at most $2t$ nodes to $C$ to obtain a $(2,2)$-CDS.
\begin{algorithm}
\caption{\textbf{Greedy Algorithm for approximate (2,2)-MCDS : Phase 1}}
\begin{algorithmic}[1]
\State Set $C = \emptyset$
\While { $\Delta_yf(C) > 0$}
\State Pick a vertex $v \in V - C$ that causes the maximum reduction in  $\Delta_yf(C)$
\State Set $C  = C \cup \{y\}$
\EndWhile
\State return $C$
\end{algorithmic}
\label{GreedyPhaseOne}
\end{algorithm}

\begin{thm}
The greedy algorithm with potential function $f$ for $(2,2)$-CDS is bounded by
the approximation ratio $(3+\ln(\Delta+2))$, where $\Delta$ is the maximum
degree of $G$.
\end{thm}

\noindent\textbf{Proof:} Assume $|V|=n.$ Let $C_{G}=\{x_{1},x_{2},\dots
{.}x_{g}\}$, in the order of nodes selected by the algorithm (phase-I). For
$0\leq i\leq g$, let $C_{i}=\{x_{1},\dots{.},x_{i}\}$. In particular, $C_{g}$
is the output of the algorithm. Suppose $C^{\ast}$ is a minimum $(2,2)$-CDS
with $\theta=|C^{\ast}|$. Since $G[C^{\ast}]$ is biconnected, we can arrange
the elements $C^{\ast}$ as $y_{1},\dots{.},y_{\theta}$ such that for each
$j\geq2$, $C_{j-1}^{\ast}=\{y_{1},\dots{.},y_{j-1}\}$ can be written as a
\textquotedblleft path", such that $y_{j}$ is connected to $y_{j-1}$, and
perhaps to the first node (or meta-node) of this path. If $y_{j}$ is also
connected to the first \textquotedblleft node", then $G[{y_{1},\dots{.},y_{j}%
}]$ is biconnected, and considered as a single meta-node. Let $C_{0}%
=C_{0}^{\ast}=\emptyset$. Since $f(C^{\ast})=2$, we have
\[
\label{potFunc2}%
\begin{split}
f(C_{i-1})-2  &  =f(C_{i-1})-f(C_{i-1}\cup C^{\ast})\\
&  =\sum_{j=1}^{\theta}\Delta_{y_{j}}(C_{i-1}\cup C_{j-1}^{\ast})\\
&  \leq\sum_{j=1}^{\theta}\left(  \Delta_{y_{j}}(C_{i-1})+1\right)
\end{split}
\]
By the pigeonhole principle, there exists a node $y_{j}$ in $C^{\ast}$ $such$
that
\[
\label{potFunc3}\Delta_{y_{j}}f(C_{i-1})+1\geq\frac{f(C_{i-1})-2}{\theta}%
\]
Since phase-I follows greedy strategy,
\[
\label{potFunc4}\Delta_{x_{i}}f(C_{i-1})\geq\Delta_{y}f(C_{i-1})\geq
\frac{f(C_{i-1})-2}{\theta}-1
\]
or
\[
\label{potFunc5}\text{or }f(C_{i})\leq f(C_{i-1})-\frac{f(C_{i-1})-2}{\theta
}+1
\]
Denote $a_{i}=f(C_{i})-2$. Then, we can equivalently write
\begin{equation}
a_{i}\leq a_{i-1}-\frac{a_{i-1}}{\theta}+1 \label{potFunc6}%
\end{equation}
Since all $a_{i}$'s are integers, we have
\[
\label{potFunc7}a_{i}\leq a_{i-1}-\left\lceil \frac{a_{i-1}}{\theta
}\right\rceil +1
\]
Now $a_{i}>\theta$ implies $\left\lceil \frac{a_{i-1}}{\theta}\right\rceil
\geq2,$ which means $a_{i}<a_{i-1}$. So long as $a_{i}>2+\theta$, phase-I
continues. Now,we can write inequality~\ref{potFunc6} as:
\[
\label{potFunc9}a_{i}\leq a_{i-1}\left(  1-\frac{1}{\theta}\right)  +1\text{,
whose solution, as
in~\cite{Guha_Khuller_1998,Ruan_Du_Jia_Wu_Li_Ko_2004,Zhou_Zhang_Wu_Xing_2014},
is }%
\]%
\[
\label{potFunc10}a_{i}\leq a_{0}\left(  1-\frac{1}{\theta}\right)  ^{i}%
+\sum_{j=0}^{i-1}\left(  1-\frac{1}{\theta}\right)  ^{j}%
\]
So after $\theta\ln(a_{0}/\theta)$ iteration, as in~\cite{Guha_Khuller_1998},
$a_{i}<2\theta$. Since phase-I continues as long as $a_{i}>\theta$, after at
most $\theta$ iterations $a_{i}\leq\theta$ since each iteration of phase-I
reduces $a_{i}^{\prime}$ by at least one unit. Suppose phase-I ends at this
stage. At this stage $f(C_{i})\leq\theta+2.$ Thus $C$ has at most $\theta+2$
biconnected components, and needs at most $2\theta+4$ additional nodes in $C$
to obtain a $(2,2)$-CDS, resulting in a bound of
\[
\label{potFunc11}\theta\ln(a_{0}/\theta)+\theta+2\theta+4=\theta\left[
\ln\frac{a_{0}}{\theta}+3+\frac{4}{\theta}\right]
\]
Asymptotically, $4/\theta$ can be ignored. So the asymptotic approximation
factor is $3+\ln(\frac{a_{0}}{\theta})=3+\ln(\frac{2n}{\theta})$. \ To bound
$\frac{2n}{\theta},$ we proceed as follows.\bigskip

Taking $i=1$, $C_{0}=\emptyset$. Then $\hat{p}(\emptyset)=0$, $q(\emptyset
)=n$, $m(\emptyset)=n$. So $f(\emptyset)=2n$. $f(x_{1})=\hat{p}(\{x_{1}%
\})+q(\{x_{1}\})+m(\{x_{1}\})$, $\hat{p}(\{x_{1}\})=0$, $q(\{x_{1}%
\})=n-|N_{G}(x_{1})|-1$, $m(\{x_{1}\})=n-1$. This implies that $f(x_{1}%
)=2n-2-|N_{G}(x_{1})|=2n-2-\Delta$. Hence
\[
\label{potFunc12}\Delta_{x_{1}}f(\emptyset)=|N_{G}(x_{1})|+2=\Delta+2
\]
Now
\begin{align}
&f(C_{1})   \leq f(C_{0})-\frac{f(C_{0})-2}{\theta}+1 ~\text{or}\\
&\frac{2n-2}{\theta}    \leq\Delta+2 ~  \text{or}\\
&\frac{2n}{\theta}   \leq\Delta+2+\frac{2}{\theta}%
\end{align}
So the approximation ratio asymptotically becomes $3+\ln(\Delta+2)$.

\section{Conclusion}

In this paper, we proposed a $(3+\ln(\Delta+2))$-approximation algorithm for
the $(2,2)$-connected dominating set for a general graph. This algorithm can
easily be generalized for the $(2,m)$-CDS problem, for $m\geq2$, resulting in
a $(3+\ln(\Delta+m))$-approximation algorithm.


\end{document}